\documentclass[prb,twocolumn,aps,amsmath,amssymb,showpacs,a4paper,10pt]{revtex4-1}
\usepackage{latexsym}
\usepackage[dvips]{graphicx}
\usepackage{pst-circ}
\usepackage[utf8]{inputenc}
\usepackage{epsfig}
\usepackage{indentfirst}
\usepackage[normalem]{ulem}
\usepackage[sort&compress]{natbib}
\usepackage[english]{babel}

\begin{document}

%%%%%%%%%%%%%%%%%%%%%%%%%%%%%%%%%%%%%%%%%%%%%%%%%%%%%%%%%%%%%%%%%%%%%%%%%%%%%%%%%%%%%%%%%%%%%%%%%%%%%%%%%%%%%%%%%%%%
%%%%%%%%%%%%%%%%%%%%%%%%%%%%%%%%%%%%%%%%%%%%%%%%%%%%%%%%%%%%%%%%%%%%%%%%%%%%%%%%%%%%%%%%%%%%%%%%%%%%%%%%%%%%%%%%%%%%%
%%%%%%%%%%%%%%%%%%%%%%%%%%%%%%%%%%%%%%%%%%%%%%%%%%%%%%%%%%%%%%%%%%%%%%%%%%%%%%%%%%%%%%%%%%%%%%%%%%%%%%%%%%%%%%%%%%%%%

\title[Phase-sticking in one-dimensional Josephson Junction Chains]{Phase-sticking in one-dimensional Josephson Junction Chains}

\author{Adem Erg\"ul, David Schaeffer, Magnus Lindblom, David B. Haviland}
 \affiliation{Nanostructure Physics, Royal Institute of Technology, SE-106 91 Stockholm, Sweden}

 \author{Jack Lidmar}
 \affiliation{Theoretical Physics, Royal Institute of Technology, SE-106 91 Stockholm, Sweden}

 \author{Jan Johansson}
 \affiliation{Department of Natural Sciences, University of Agder, Kristiansand, Norway}

\begin{abstract}

We studied current-voltage characteristics of long one dimensional Josephson junction chains with Josephson energy much larger than charging energy, $E_J \gg E_C$. In this regime, typical IV curves of the samples consist of a supercurrent branch at low bias voltages followed by a voltage-independent chain current branch, $I_{Chain}$ at high bias. Our experiments showed that $I_{Chain}$ is not only voltage-independent but it is also practically temperature-independent up to $T_C$. We have successfully model the transport properties in these chains using a capacitively shunted junction model with nonlinear damping.

\end{abstract}
\maketitle

%%%%%%%%%%%%%%%%%%%%%%%%%%%%%%%%%%%%%%%%%%%%%%%%%%%%%%%%%%%%%%%%%%%%%%%%%%%%%%%%%%%%%%%%%%%%%%%%%%%%%%%%%%%%%%%%%%%%%
\section{Introduction}
\indent  Josephson Junction (JJ) chains exhibit many interesting phenomena such as Coulomb blockade of Cooper pairs \cite{Agren}, coherent phase-slips \cite{Haviland}, synchronous Cooper pair tunneling \cite{Andersson1} and  superinsulation \cite{Vikonur}. These properties are utilized for various applications such as the development of the Fluxonion for quantum information precessing \cite{Manucharyan}, development of voltage standards in metrology \cite{Kautz} and for widely tunable parametric amplifiers \cite{Castellanos}. Furthermore it is suggested that very long one-dimensional Josephson junction chains formed in a transmission line geometry can be employed for creating an analog of the event horizon and Hawking radiation \cite{Nation,Nation2}.

\indent The Josephson junction is described by two ratios: the ratio of the characteristic energies, the Josephson energy ($E_J$) and the charging energy ($E_C$), and the ratio of the effective damping resistance  ($R_{damp}$) and the quantum resistance ($R_Q=h/4e^2=6.45k\Omega$). Depending on these ratios, either the charge or the phase  behaves as a classical variable. There have been extensive studies of long and compact chains of Josephson Junctions in the limit $E_J/E_C\ll 1$ and  $R_Q/R_{damp}\ll 1$ \cite{Watanabe, Haviland2,Andersson2,Agren}. In this extreme, the JJ chain forms a high impedance transmission line \cite{Agren} for the Josephson plasmon mode \cite{Mooij} and when this impedance exceeds the quantum resistance $R_Q$, coherent quantum phase slips \cite{Nazarov} give rise to a Coulomb blockade of Cooper pair tunneling \cite{Averin}. This phenomena is the quantum mechanical complement of the Josephson Effect. While numerous groups have observed the Coulomb blockade of Cooper pair tunneling \cite{Maibaum,Schafer,vanderzant} a robust demonstration of the complement to the AC Josephson effect, or synchronization to Bloch oscillations, is yet to be demonstrated.

\indent In the other extreme, $E_J/E_C\gg 1$ and $R_Q/R_{damp}\gg 1$, the phase of the junction can be treated as a classical variable while the charge fluctuates strongly. Recently it was demonstrated that chains in this regime can be used as so called superinductors which have a high frequency impedance much larger than the quantum resistance \cite{Masluk,Bell}. There have been several successful experiments of the observation of quantum phase-slips in chains with large Josephson energy \cite{Pop2,Manucharyan2}.
 
\begin{figure}[t]
 \centering
 \includegraphics[width=0.45\textwidth]{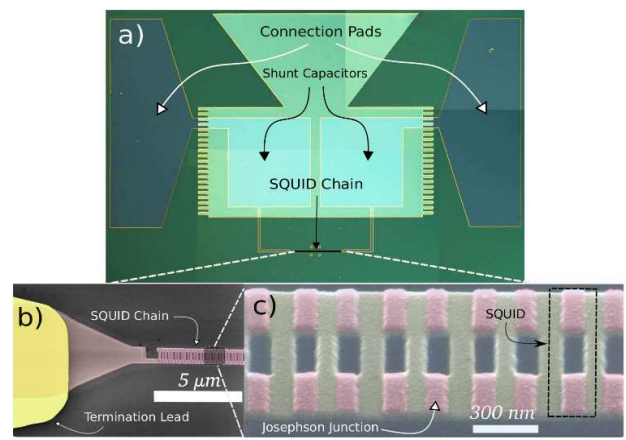}
 \caption{ a)Optical Microscope images of a Josephson Junction chain. The chain consists of 2888 SQUIDS with total length $500\mu m$. Shunt capacitors with the sizes $1000\mu m$ x $800 \mu m$ are placed to the input and output of the chain. b) Scanning Electron Microscope (SEM) images of SQUID chains together with termination lead. c) A group of SQUIDs which  consists of two parallel Josephson junctions with the dimensions $300nm$ x $100 nm$.}
 \label{fig:fig1}
\end{figure} 
    
\indent The aim of this study is to understand the current-voltage characteristics of long Josephson junction chains. We have fabricated and studied one dimensional Josephson junction chains of three different lengths (384, 2888, and 4888 junctions) in the regime where the Josephson energy is much larger than the charging energy, $E_J\gg E_C$. We characterize the damping by the normal state resistance of the junctions, $R_N$ which falls in the range, $0.01\leq R_Q/R_N \leq 1$. The DC IV curves of these samples consists of a super-current, S.C., branch at low bias voltages followed by a voltage-independent chain current branch, $I_{Chain}$ which happens to be only a small fraction of the the Ambegaokar-Baratoff critical current for a single junction, $I_C$ \cite{Ambegaokar}, $I_{Chain}/I_C \sim 0.2$.  We focus on the large voltage behavior, $2\Delta_0/e \ll V < N2\Delta_0/e$, where the classical phase-slips \cite{Langer,McCumber} are determining factor for the phase dynamics in the chain. We did simulations in order to understand the complicated phase-slip dynamics which occur inside the Josephson junction chain causing this novel behavior.

\indent The paper is organized as follows, in Sec. II we describe the fabrication process and the measurement setup. In Sec. III we have presented experimental results together with a circuit model and details of the simulations. And in the final section we present the conclusions.   

%%%%%%%%%%%%%%%%%%%%%%%%%%%%%%%%%%%%%%%%%%%%%%%%%%%%%%%%%%%%%%%%%%%%%%%%%%%%%%%%%%%%%%%%%%%%%%%%%%%%%%%%%%%%%%%%%%%%%
\section{Experimental}

\begin{figure}[t]
\centering
\includegraphics[width=0.43\textwidth]{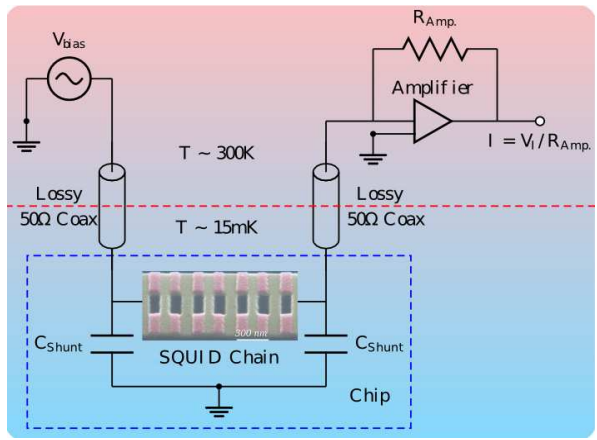}
\caption{Schematic diagram of the measurement circuit.}
\label{fig:fig2}
\end{figure}

\indent The Josephson Junction chains consist of serially connected SQUIDs (Superconducting QUantum Interference Devices). A SQUID is formed by connecting two Josephson Junctions in parallel and when the  SQUIDs loop inductance is small compared to the Josephson inductance, $L_{loop}\ll L_J$, where $L_J=\hbar /(2eI_C)$ (which is the case in our samples), the effective Josephson coupling energy, $E_J$ of Josephson Junctions can be modulated  by an external magnetic field, $E_J=E_{J0}|\cos(2\pi \Phi_{ext}/\Phi_0)|$. 

\indent Fig.~\ref{fig:fig1}(a) shows the optical microscope image of a sample together with shunt capacitors and connection pads. Two thin film capacitors are fabricated on-chip and connected in series to shunt the Josephson junction chain ($C_{Shunt}\sim 1nF$) in order to reduce the high frequency impedance seen by the chain  and provide filtering for the fluctuations coming from the external leads and circuitry. The first layer of the shunt capacitors are Al rectangles defined by optical lithography on $Si/SiO_2$ substrate. The insulating layer is formed by sputtering a $15nm$ thick $SiO_2$. The final layer is formed by depositing  Au connection pads, there by creating $Al/SiO_2/Au$ capacitors. Fig.~\ref{fig:fig1}(b)-(c) shows scanning electron microscope images of a chain together with a termination lead. The Josephson junction chain is defined by Electron Beam Lithography and the overlapping $Al/Al_2O_3/Al$ tunnel junctions are made by the standard double angle shadow evaporation technique \cite{Dolan}.

\indent All the experiments are conducted in a dilution refrigerator with a base temperature of $\sim15~mK$. Fig.~\ref{fig:fig2} shows the schematic diagram of the measurement circuit. The sample is mounted on a printed circuit board which is in turn mounted in a RF tight copper can. The measurement leads in the fridge are made of lossy $50\Omega$ coax cables.

%%%%%%%%%%%%%%%%%%%%%%%%%%%%%%%%%%%%%%%%%%%%%%%%%%%%%%%%%%%%%%%%%%%%%%%%%%%%%%%%%%%%%%%%%%%%%%%%%%%%%%%%%%%%%%%%%%%%%
\section{Results}
\begin{figure}[t]
  \centering
  \framebox{
  \includegraphics[width=0.4\textwidth]{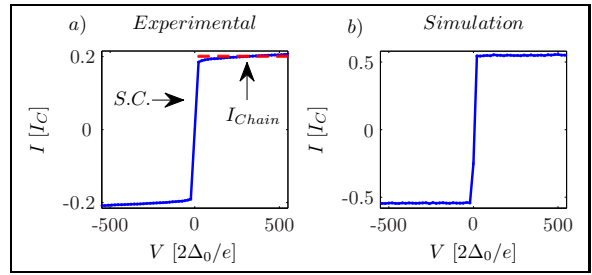} 
  }
  \caption{ Experimental(a) and simulated(b) DC IV curves of a long Josephson Junction chain with $N=2888$ SQUIDS, $E_J/E_C=36$ at zero magnetic field and $R_Q/R_N=0.08$. $I_C=634~nA$ is the Ambegaokar-Baratoff critical current for a single junction in this chain, \cite{Ambegaokar}, and superconducting energy gap of $Al$ is $\Delta_0=200\mu eV$.}
\label{fig:fig3}
\end{figure}

%%%%%%%%%%%%%%%%%%%%%%%%%%%%%%%%%%%%%%%%%%%%%%%%%%%%%%%%%%%%%%%%%%%%%%%%%%%%%%%%%%%%%%%%%%%%%%%%%%%%%%%%%%%%%%%%%%%%%
\subsection{Voltage Independent Chain Current}

\indent Typical DC IV curves of long Josephson Junction chains with $E_J/E_C\gg1$ and $R_Q/R_N\gg 1$ show practically voltage-independent constant current branches, $I_{Chain}$ between the supercurrent branch and the normal tunneling branch. As an example, experimental and simulated DC IV curves of a long Josephson Junction chain with $E_J/E_C=36$ in zero magnetic field and $R_Q/R_N=0.08$ are shown in Fig.~\ref{fig:fig3}. The experimental DC IV curve consists of a supercurrent, S.C., branch at low bias voltages and a voltage-independent constant current branch at higher bias voltages. This sample had 2888 junctions in series. The total junction area of a single SQUID is $A_{\mathrm{Jun}}=0.05\mu m^2$ and the normal state resistance of a single SQUID is $R_N = 0.5 k\Omega$. The charging energy is defined by $E_C=e^2/(2C_{S}A_{\mathrm{Jun}})$ with $C_S=45 fF/\mu m^2$ being the specific capacitance. Throughout the paper the quoted critical current value is the low temperature limit of the calculated Ambegaokar-Baratoff critical current for the single SQUID, $I_C=\pi \Delta_0/(2eR_N)$ \cite{Ambegaokar}.

\indent The simulated DC IV curve of the sample is shown in Fig.~\ref{fig:fig3}(b) and the details of the simulations will be given below. It is important to emphasize that the simulation parameters were either experimentally measured or estimated from sample geometry. We find qualitative agreement between the simulation and experiment and by adjusting only the critical current $I_C$, it is possible to get quantitative agreement between experimental and simulated DC IV curves. We therefore conclude that the circuit model accurately simulates the phase-dynamics of the Josephson junction chains.

\begin{figure}[t]
  \centering
  \framebox{
    \includegraphics[width=0.4\textwidth]{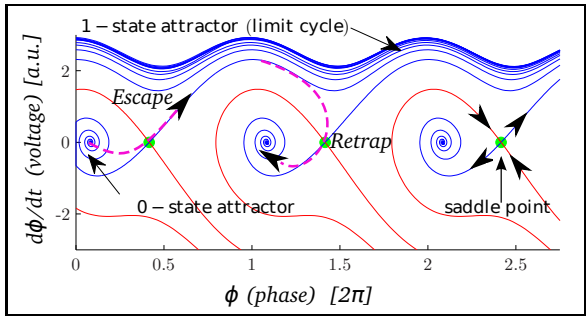}
    }
  \caption{Phase-space diagram of a single junction from the RCSJ model with $Q=5$ and $I=0.5I_C$.}
\label{fig:fig4}
\end{figure}

\indent In order to gain insight into the complex dynamics of the Josephson junction chains, we appeal to the analogy with a simpler and more well-studied Resistively Capacitively Shunted Junction (RCSJ) model \cite{Likharevbook}. In normalized units this model gives the current as
\begin{equation}
I/I_C=\ddot{\phi} + \dot{\phi}/Q+\sin(\phi), 
\label{eq:ITotIC}
\end{equation}
where $\phi$ is the phase difference over the junction. The damping in this model is due to a frequency-independent ohmic shunt resistance, $R$,  and it is expressed in terms of a dimensionless quality factor $Q = \pi^2(R/R_Q)^2(E_J/2E_C)$. This quality factor is sometimes called the Stewart-McCumber damping parameter $\beta=Q^2$. 

\indent  A very well known mechanical analog for the RSCJ model is a particle in a tilted washboard potential. In the mechanical model $1/Q$ corresponds to friction and hence small $Q$ represents large damping. This model has two distinct states, one is the  particle resting at the potential minimum (0-state) corresponding to the S.C. branch and the other state is the particle running down the washboard potential (1-state) corresponding to the dissipative branch. A graphical way to visualize the dynamics of a Josephson junction is a phase-space diagram. Fig.~\ref{fig:fig4} shows the phase-space diagram of a Josephson junction with underdamped dynamics biased below the critical current ($Q=5$ and $I=0.5I_C$).  Two basins of attraction are separated by the red lines.  A particle escaping from the 0-state through the saddle point (green dot, indicating the local maximum of the tilted washboard potential) can not move directly to the basin of the next 0-state attractor without entering the basin of 1-state attractor.  For fluctuation-free dynamics, once the particle is at the 1-state it will continue to run down the washboard potential.

\indent Fluctuations, which necessarily accompany the damping, give rise to transitions between the two stable attractors, known as escape and retrapping.  The energy required to switch the system from 1-state to 0-state is called {\em activation~energy} and it is approximately equal to the kinetic energy of the particle in the 1-state. The kinetic energy decreases with increasing damping. Therefore unstable switching between the 0-state and 1-state require large thermal energy and/or large damping. In Refs. \cite{Martinis,Kautz2} it was shown that in a certain range of parameters, the RCSJ model with noise current predicts that both of these states can be unstable and junction can switch rapidly back and forth between running and resting states, creating a constant current branch on DC IV curve. While these simulations were performed on a single junction with linear damping, we propose that this type of instability leads to a voltage-bias independent current due to a continuous phase-slipping and phase-sticking in Josephson junction chains.

%%%%%%%%%%%%%%%%%%%%%%%%%%%%%%%%%%%%%%%%%%%%%%%%%%%%%%%%%%%%%%%%%%%%%%%%%%%%%%%%%%%%%%%%%%%%%%%%%%%%%%%%%%%%%%%%%%%%%
\subsection{Simulations}

\begin{figure}[t]
\centering
\framebox{
\includegraphics[width=0.48\textwidth]{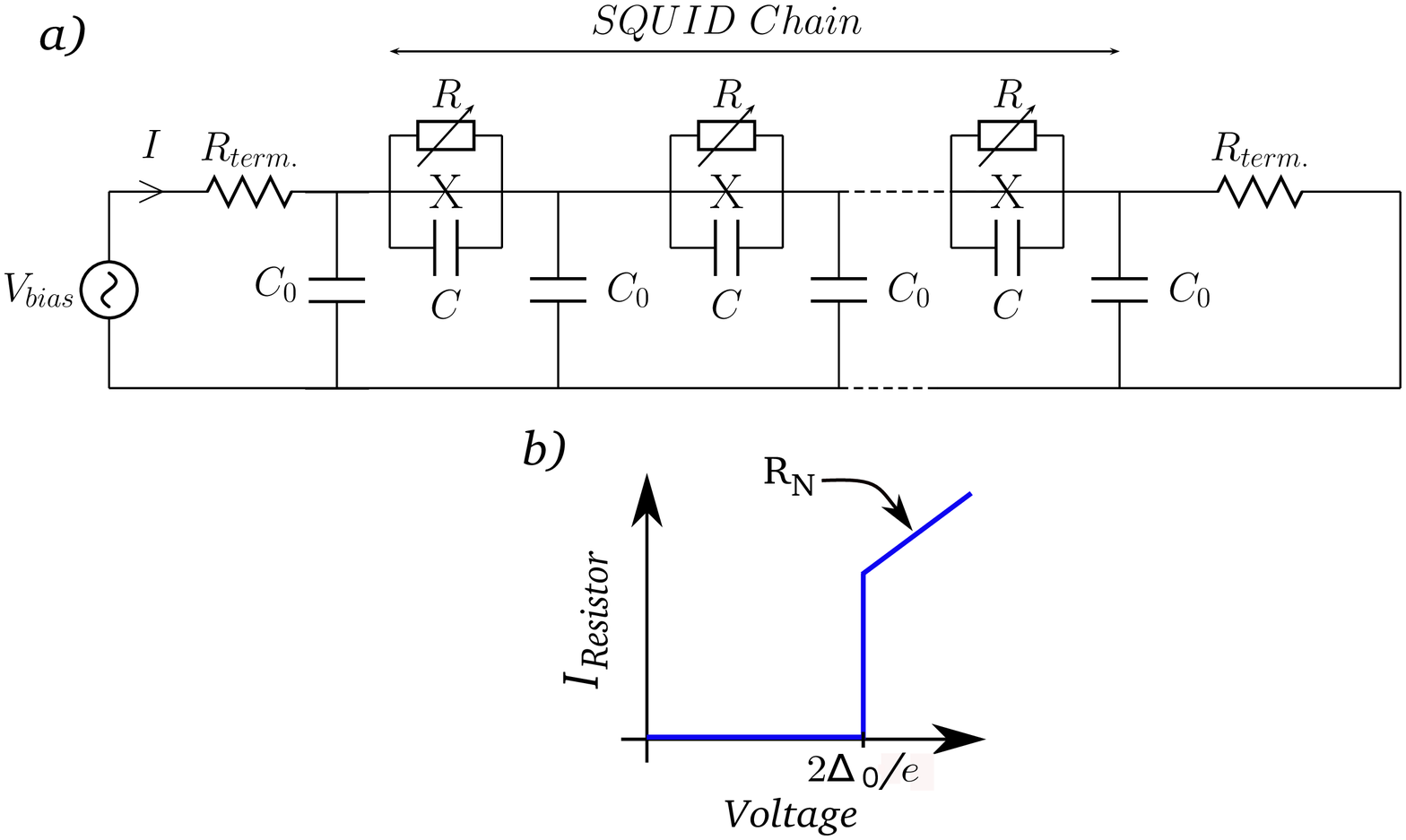}
}
\caption{a) A circuit model of the $SQUID$ Chain. b) Nonlinear resistive current through a junction in the chain.}
\label{fig:fig5}
\end{figure}

\indent The dynamics of a Josephson junction chain is far more complex than that of the simple RCSJ model. In the chain, collective modes can exist and the damping is far more complicated than a simple ohmic resistor. To address this we performed simulations of long chains, modeled using the circuit diagram shown in Fig.~\ref{fig:fig5}(a). Each junction of the SQUIDs in the chain is modeled as an ideal Josephson junction shunted by a capacitance $C$ and a nonlinear resistor $R$, which only lets current through when the voltage across it exceeds the gap voltage $V_g=2\Delta_0/e$. The total current through junction $i$ is thus
\begin{align}
\label{eq:1}
  I^\mathrm{tot}_i &= I^s_i + I^C_i + I^R_i \nonumber \\
&= 
I_c \sin(\theta_{i} - \theta_{i+1}) + C (\dot V_{i} - \dot V_{i+1}) + I^R_i ,
\end{align}
where $\theta_i$ is the phase of the superconducting order parameter at the island to the left of junction $i$, and $V_i = \hbar \dot \theta_i / 2e$ is the voltage. The nonlinear resistive current is taken to be
\begin{equation}
  I^R_i =
  \begin{cases}
    (V_{i} - V_{i+1})/R + I^n_i & \text{if}\,  |V_{i} - V_{i+1}| > V_g \\
    0 & \text{otherwise}
  \end{cases}
\label{eq:nonlinear}
\end{equation}
as shown in Fig.~\ref{fig:fig5}(b), where $R$ is the normal resistance of a single junction.  (The sub-gap resistance is thus assumed to be infinite.) In addition a thermal noise current $I^n$ is included in Eq.~(\ref{eq:nonlinear}). The latter is modeled as a Gaussian random Johnson-Nyquist noise with zero mean and covariance $\left< I^n_i(t) I^n_j(t')\right> = (2 k_B T/R) \delta_{ij} \delta(t-t')$. Experimentally, the Josephson junction chain is voltage biased. Therefore, the currents entering the chain from the left through the left lead resistance and leaving the chain on the right are given by  
\begin{equation}
  \label{eq:3}
  I_L = (V_{bias} - V_1)/R_\text{term} + I^n_L ,
  \qquad
  I_R = V_N/R_\text{term} + I^n_R
\end{equation}
The Johnson-Nyquist noise $I^n_{R,L}$ in the terminal resistors have zero mean and obey $\left< I^n(t)  I^n(t')\right> = (2 k_B T/R_\text{term}) \delta(t-t')$. These terminal resistances consists of lead resistances together with the characteristic impedance of the coaxial cables, approximately equal to $Z_0/2\pi \approx 60\Omega$, where $Z_0$ is the free space impedance. In our simulations we therefore set $R_\text{term} = 50\Omega$. This low impedance is a main source of dissipation and noise in the system. Now, Kirchhoff's law holds at each superconducting island,
\begin{equation}
  \label{eq:4}
  C_0 \dot V_i + I^\text{tot}_{i} - I^\text{tot}_{i-1} = 0,
\end{equation}
where $C_0$ is the capacitance to ground. This gives a coupled system of 2nd order differential equations for the superconducting phases $\theta_i$. These are integrated with a symmetric time discretization using a leap-frog scheme, with a small time step $\Delta t = 0.02 (\hbar / 2e I_c R) = 0.02 (RC/Q^2)$. Each iteration requires the solution of a tridiagonal system of equations. By varying the bias voltage and calculating the resulting current we obtain the IV-characteristics of the structure. The voltage is stepped up slowly from zero, or down from a high value, to avoid sharp transient effects near the left lead where the voltage is applied. We also keep track of the locations and times of phase slip events, i.e. when the phase difference across a junction $\theta_{i}-\theta_{i+1}$ passes between the disjoint intervals $I_m = [-\pi + 2\pi m, +\pi + 2\pi m]$ for integer $m$.

%%%%%%%%%%%%%%%%%%%%%%%%%%%%%%%%%%%%%%%%%%%%%%%%%%%%%%%%%%%%%%%%%%%%%%%%%%%%%%%%%%%%%%%%%%%%%%%%%%%%%%%%%%%%%%%%%%%%%
\subsection{$I_{Chain}/I_C$ as a function of $E_J/E_C$}

\begin{figure}[t]
  \centering
  \framebox{  \includegraphics[width=0.28\textwidth]{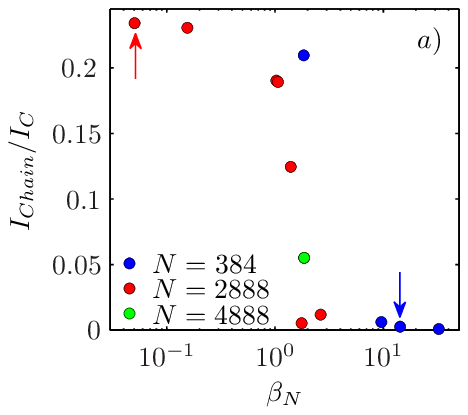}
              \includegraphics[width=0.2\textwidth]{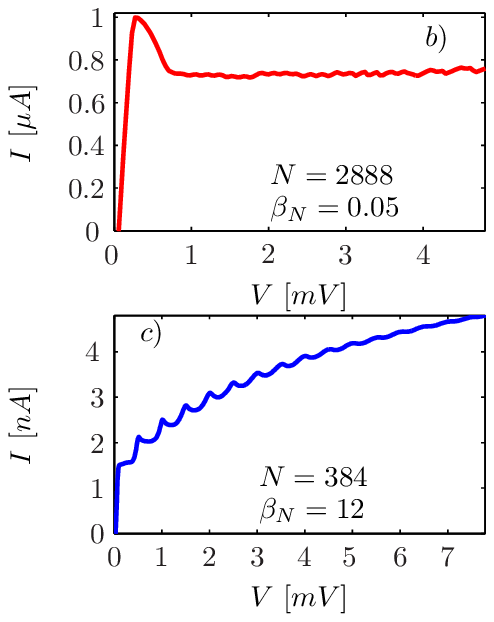}}
              
  \caption{a)$I_{Chain}/I_C$ versus $\beta_N=Q^2$ for various samples. Each data point represents the measurement of a different sample at zero magnetic field. The IV curves in (b)and (c) are of the samples marked with the arrows.}
\label{fig:fig6}
\end{figure}

\indent Fig.~\ref{fig:fig6}-a shows the $I_{Chain}/I_C$ as a function of $\beta_N$ for various samples.  Each data point represents the measurement of a different sample at zero magnetic field. The data in the figure is collected from the measurements of $12$ different samples with three lengths and different critical currents. The chain current, $I_{Chain}$, is taken as the voltage-independent current at large bias voltages ($V\sim 100\Delta_0/e$). The figure consists of two distinct parts which are separated by a rapid decrease of the $I_{Chain}/I_C$ level around $\beta_N\sim 1$ where dynamics of the long Josephson junction chains undergo a qualitative change. This behavior is consistent with the what we expect from the RCSJ model. At $\beta_N\ll 1$ the damping is strong and activation energies for escape and retrapping are similar, every phase-slip event (escape) is followed by a phase-sticking (retrapping). As will be discussed below, this behavior is consistent with the phase-slips happening randomly throughout the chain and a continuous slip-stick process is manifest as a constant current branch in the DC IV curves  (fig.~\ref{fig:fig6}-b). The rate of phase-slip and phase-stick is determined by the bias voltage. In the opposite limit when $\beta_N\gg 1$ the damping is small which inhibits phase-sticking. Once the junction starts slipping it continues to slip, resulting in lower $I_{Chain}/I_C$ and features in the DC IV curve at multiples of the gap voltages, $V=n2\Delta_0/e$ where the nonlinear damping rapidly increases (fig.~\ref{fig:fig6}-c). 

%%%%%%%%%%%%%%%%%%%%%%%%%%%%%%%%%%%%%%%%%%%%%%%%%%%%%%%%%%%%%%%%%%%%%%%%%%%%%%%%%%%%%%%%%%%%%%%%%%%%%%%%%%%%%%%%%%%%%
\subsection{Simulated Results on the Effect of Damping}
\begin{figure}[t]
  \centering
  \framebox{
  \includegraphics[width=0.47\textwidth]{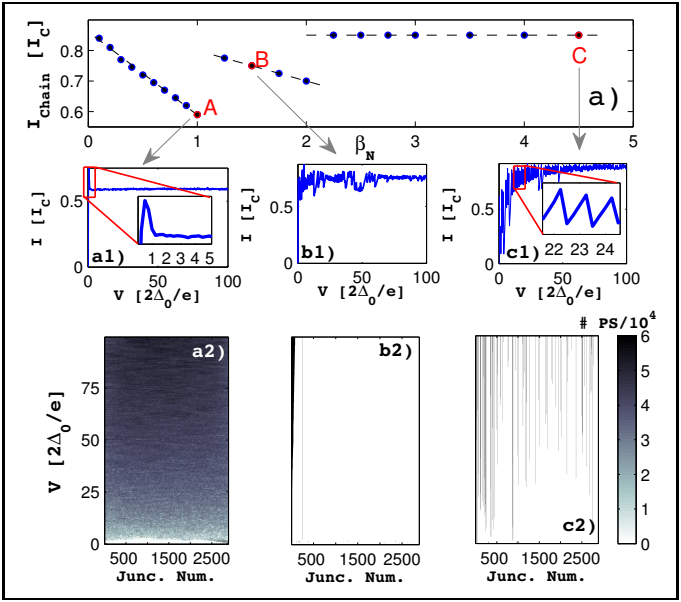}
   }
  
  \caption{ a)$I_{Chain}$ as a function of $\beta_{N}$. DC IV curves (a1,b1,c1) and contour plots of the phase-slips distributions (a2,b2,c2) across the Josephson junction chains for three different $\beta_{N}$ values.} 
\label{fig:fig7}
\end{figure}

\indent In this section we present simulation results. The experimental parameters used in the simulations are $k_BT/E_J=0.0033$, $C_0/C=0.01$, and the damping parameter is tuned between $\beta_{N}=0.1$ and $\beta_{N}=4.5$. Fig.~\ref{fig:fig7} (a) shows the $I_{Chain}/I_C$ as a function of $\beta_{N}$ and there are three distinct branches. These branches correspond to different phase-slips distributions in the chains. We have selected one point from each branch, $~A,~B$ and $C$ ($\beta_{N}=1$, $\beta_{N}=1.5$ and $\beta_{N}=4.5$), and plotted the DC IV curves together with a contour plot of the phase-slip distribution across the chain. We emphasize that all points on the different branches A-C have qualitatively similar phase-slip distributions and current-voltage characteristics. The DC IV curve of point A with $\beta_{N}=1.0$  is shown at Fig.~\ref{fig:fig7}(a1) and it consists of a supercurrent branch followed by a uniform current level very similar to the IV curve shown in Fig.~\ref{fig:fig6}(b).  Fig.~\ref{fig:fig7}(a2) shows the contour plot of the phase-slip distribution across the chain as a function of bias voltage and this distribution is consistent with the picture where every phase-slip is followed by a phase-stick. This behavior is expected in the high damping regime and phase-slips happens randomly throughout the chain, without preference for any particular point in space.

\indent Fig.~\ref{fig:fig7}(c1) shows the DC IV curve of point C with $\beta_{N}=4.5$. There are strong gap features in the IV curve and the dissipative branch flattens out at large bias voltages. The phase-slip distribution shows that once a junction starts slipping it continues this motion without re-trapping and these slipping junctions are randomly distributed across the chain (Fig.~\ref{fig:fig7}-c2). This simulation point corresponds to the IV curve in Fig.~\ref{fig:fig6}(c) where $\beta_N\gg 1$. And finally, Fig.~\ref{fig:fig7}(b1) shows the DC IV curve of point B with $\beta_{N}=1.5$. Here we also see a S.C. branch followed by a constant current branch similar to the point A. But there are strong fluctuations on the constant current branch which are not present at point A. These fluctuations are very similar to the gap features seen in the IV curves for point C. There is one specific junction at one end of the chain where the phase-slip nucleates and as the bias voltage is increased neighboring junctions starts slipping one by one.

\begin{figure}[t]
  \centering
  \framebox{
  \includegraphics[width=0.2\textwidth]{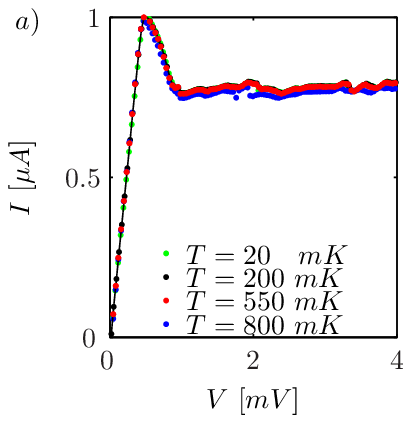}
  \includegraphics[width=0.2\textwidth]{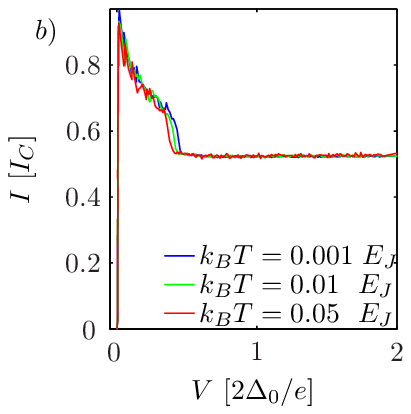}
  }
  \caption{ a) IV curves of a sample ($N=2888$ and $\beta_N\sim 1$) at zero magnetic field with various temperatures. b) Simulated IV curves with the same sample parameters.}
  \label{fig:fig8}
\end{figure}

\indent Fig.~\ref{fig:fig8}(a) shows the DC IV experimental curves of a sample for the temperatures between $20mK$ up to $800mK$, at zero magnetic field. Fig.~\ref{fig:fig8}-b shows the simulated IV curves with the experimental parameters of the sample and similar range of temperatures, $k_BT/E_J=0.001~to~0.05$. Experimental results and the simulations are in good agreement and both show that $I_{Chain}$ is independent of temperature. 

This temperature independence is an indication that the overall current voltage characteristic is determined primarily by the damping and not driven by fluctuations. The finite slope of the S.C. branch observed in the experiment (Fig.~\ref{fig:fig8}-a) is probably due to quantum phase slips which are not accounted for in the classical simulation. The simulation (Fig.~\ref{fig:fig8}-b) does take into account phase-diffusion  or thermally activated phase slips on S.C. branch but these are not able to account for the observed slope. The simulation also nicely reproduces the observed peak in current at low bias voltages.

%%%%%%%%%%%%%%%%%%%%%%%%%%%%%%%%%%%%%%%%%%%%%%%%%%%%%%%%%%%%%%%%%%%%%%%%%%%%%%%%%%%%%%%%%%%%%%%%%%%%%%%%%%%%%%%%%%%%%
\subsection{High Voltage Characteristics}

\begin{figure}[t]
  \centering
  \framebox{ 
  \includegraphics[width=0.4\textwidth]{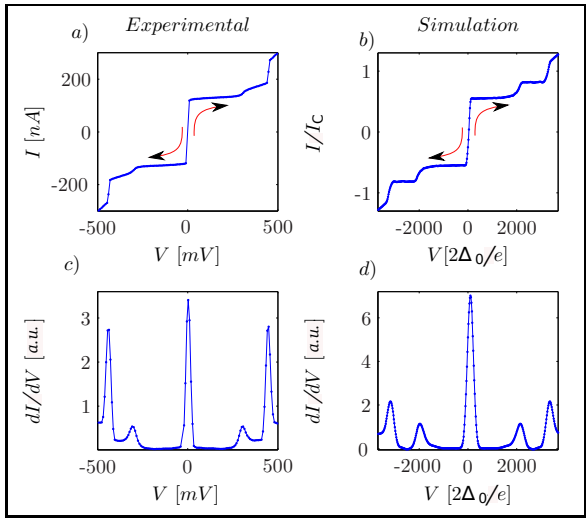} 
  }
  \caption{a)High-voltage characteristics of the sample shown in Fig.~\ref{fig:fig3}. Differential conductance, $dI/dV$ (b), simulated DC IV curve (c) and the simulated $dI/dV$ (d) of the same sample. The arrows shows the direction of the sweep.}
\label{fig:fig9}
\end{figure}

\indent Fig.~\ref{fig:fig9} shows the high-voltage characteristics of the same sample as in  Fig.~\ref{fig:fig3}. The large-scale differential conductance (Fig.~\ref{fig:fig9}-c) curve shows five distinct peaks: the S.C. peak around zero bias, two peaks at $V=\pm450mV$ corresponding to the sum-gap voltage, and two additional peaks at $V=\pm300mV$. These extra peaks are due to a transition between two different chain currents in the IV curves, fig.~\ref{fig:fig9}-a). These peaks are also visible in the simulated current voltage characteristics and the overall shape of the large scale IV curve is well reproduced by the simulation.

\begin{figure}[t]
 \centering
\framebox{ 
    \includegraphics[width=0.47\textwidth]{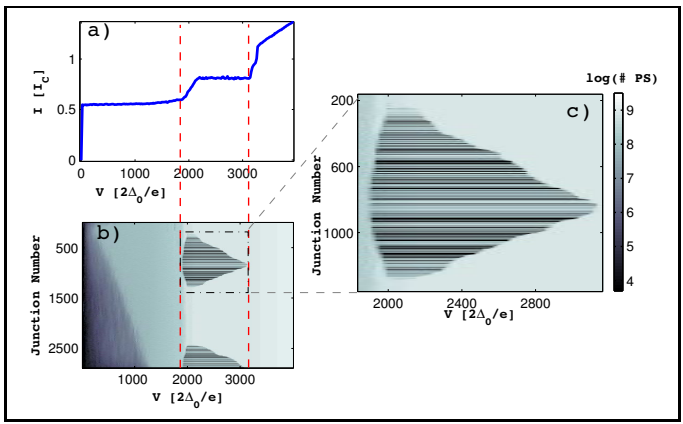} 
         }
  \caption{ Simulated DC IV curve (a) and distribution of phase-slips (b-c) for the sample shown in Figs.~\ref{fig:fig3} and ~\ref{fig:fig9}.}
\label{fig:fig10}
\end{figure}

\indent Fig.~\ref{fig:fig10}(a) shows the simulated IV curve of the sample between zero bias up to the normal tunneling branch.  Fig.~\ref{fig:fig10}(b) shows the distribution of the phase-slips across the array, the gray scale  represents the number of phase-slips at each junction and voltage. At low voltages phase-slips tend to accumulate at one end of the chain. At higher voltages they become uniformly distributed across the chain. This random phase-slip distribution is similar to the simulated behavior seen in Fig.~\ref{fig:fig7}(A1) with high damping. Surprisingly, as the bias voltage is further increased a group of junctions loose this randomness and form a cluster with a fixed number of phase-slips, independent of bias voltage (Fig.~\ref{fig:fig10}-c). 

\indent This cluster rapidly grows with increasing bias voltage, causing an increase of the chain current. At the maximum extent there are approximately, $N\sim1100$ junctions in one of the clusters, which is a considerable part of the chain. After that point as the voltage bias further increased the size of the cluster gradually decreases with the junctions leaving the cluster in the opposite order as they were added. This gradual decrease creates a second flat branch in the simulated IV curve. Thus we see that the simulation allows us to gain insight into the complex dynamics and study in detail the phase-slip distribution throughout the chain. 

%%%%%%%%%%%%%%%%%%%%%%%%%%%%%%%%%%%%%%%%%%%%%%%%%%%%%%%%%%%%%%%%%%%%%%%%%%%%%%%%%%%%%%%%%%%%%%%%%%%%%%%%%%%%%%%%%%%%%
\section{Conclusion}

\indent In this paper we have presented experimental observation of voltage independent constant current branch, $I_{Chain}$  in the IV curves of long Josephson junction chains with $\beta_N<1$. We have successfully simulated the current voltage characteristics in this regime with a coupled RCSJ model. The observation of voltage-independent chain current, $I_{Chain}$ is a manifestation of a random process of phase-slipping and phase-sticking that is uniformly distributed throughout the chain. The phase slip rate is defined by the bias voltage and our simulations showed that voltage independent constant current branch is created by uncorrelated phase slips. Moreover experimental results showed that there is a significant decrease of $I_{Chain}/I_C$ when $\beta_N>1$.

\indent Simulations showed that the damping parameter, $\beta_{N}$, is important for defining the distribution of phase-slips and phase-sticking processes in the Josephson junction chains. Different phase-slip distributions and phase-sticking processes creates various shapes of the DC IV curves and it is possible gain insight to these processes  by just analyzing the shape of the IV curves. Furthermore, our simulations showed that $I_{Chain}$ is independent of temperature up to $T_C$ and we confirmed this experimentally. Despite the fact that our model does not take into account quantum tunneling, we found good agreement between our classical model and the experimental data. In particular, the shape of the DC IV curve was determined by the distribution of phase slipping and phase sticking events.  We conclude that phase-slipping together with phase-sticking, is the dominant mechanism which defines the dynamics of the long Josephson Junction chains at finite voltages.

 %%%%%%%%%%%%%%%%%%%%%%%%%%%%%%%%%%%%%%%%%%%%%%%%%%%%%%%%%%
\def\newblock{\hskip .11em plus .33em minus .07em}

%%%%%%%%%%%%%%%%%%%%%%%%%%%%%%%%%%%%%%%%%%%%%%%%%%%%%%%%%%%%%%%%%%%%%%%%%%%%%%%%%%%%%%%%%%%%%%%%%%%%%%%%%%%%%%%%%%%%%
\end{document}